\documentclass[%
 aps,
 amsmath,amssymb,
 prab,
 reprint,%
 twocolumn,
]{revtex4-2}

\usepackage{graphicx}
\usepackage{dcolumn}
\usepackage{bm}
\usepackage{amsmath}
\usepackage[algo2e]{algorithm2e}
\usepackage{siunitx}

\usepackage[textsize=tiny,disable]{todonotes}
\usepackage{printlen}
\uselengthunit{in}
\newlength\imageheight
\newlength\imagewidth

\DeclareMathOperator*{\argmax}{arg\,max}
\DeclareMathOperator*{\argmin}{arg\,min}
\newcommand{\xX}{\mathcal{X}}
\newcommand{\dD}{\mathcal{D}}
\newcommand{\hH}{\mathcal{H}}
\newcommand{\oO}{\mathcal{O}}
\newcommand{\RR}{\mathbb{R}}
\newcommand{\NN}{\mathbb{N}}
\newcommand{\PP}{\mathbb{P}}
\newcommand{\bB}{\mathcal{B}}
\newcommand{\aA}{\mathcal{A}}
\newcommand{\sS}{\mathcal{S}}
\newcommand{\CMAES}{\textsc{CMA-ES}}
\newcommand{\NelderMead}{\textsc{NelderMead}}
\newcommand{\LineBO}{\textsc{LineBO}}
\newcommand{\CLineBO}{\textsc{C-LineBO}}
\newcommand{\CLineBOloc}{\textsc{C-LineBO-loc}}
\newcommand{\ALineBO}{\textsc{A-LineBO}}
\newcommand{\ALineBOunsafe}{\textsc{A-LineBO (unsafe)}}
\newcommand{\ALineBOloc}{\textsc{A-LineBO-loc}}

\newcommand{\eqdef}{\stackrel{\text{def}}{=}}

\begin{document}

\preprint{AIP/123-QED}

\title[Short Title]{Tuning Particle Accelerators with Safety Constraints using Bayesian Optimization}

\author{Johannes Kirschner}
\author{Mojmir Mutn\'y}
\author{Andreas Krause}
\affiliation{Department of Computer Science, ETH Zurich, 8232 Zurich, Switzerland}
\email{jkirschn@ualberta.ca}
\author{Jaime Coello de Portugal}%
\author{Nicole Hiller}%
\author{Jochem Snuverink}%
\affiliation{ 
	Paul Scherrer Institut, 5232 Villigen PSI, Switzerland
}
\email{jochem.snuverink@psi.ch}
%
%

\date{\today}

\begin{abstract}
Tuning machine parameters of particle accelerators is a repetitive and time-consuming task that is challenging to automate. While many off-the-shelf optimization algorithms are available, in practice their use is limited because most methods do not account for safety-critical constraints in each iteration, such as loss signals or step-size limitations. One notable exception is safe Bayesian optimization, which is a data-driven tuning approach for global optimization with noisy feedback. We propose and evaluate a step-size limited variant of safe Bayesian optimization on two research facilities of the Paul Scherrer Institut (PSI): a) the Swiss Free Electron Laser (SwissFEL) and b) the High-Intensity Proton Accelerator (HIPA). We report promising experimental results on both machines, tuning up to 16 parameters subject to 224 constraints.
\end{abstract}

\keywords{Bayesian optimization, particle accelerators, parameter tuning, safety constraints}
\maketitle

\section{Introduction}

Particle accelerators are complex machines consisting of many elements
and are typically built according to an idealized design. 
However, in reality, there are always systematic and time-varying errors
that will reduce the performance with respect to the design performance.
These errors need to be corrected by parameter tuning to achieve the optimal performance.
Therefore, empirical parameter tuning is a required and often reoccurring task for any particle accelerator.
Common tuning objectives include the beam shape, beam trajectory,
beam loss minimization, or a combination of multiple objectives.

In this paper, we present a method that allows for safe optimization of particle accelerators. Our approach follows recent work by \citet{kirschner2019linebo} on safe Bayesian optimization. The main idea behind Bayesian optimization is to compute regression estimates of the target and loss functions using the data collected during optimization  \cite{Mockus1982,frazier2018tutorial}. The estimates and associated statistical uncertainty are then used to systematically explore the parameter space, while guaranteeing that the query points are safe. Our main contribution is an experimental evaluation of safe Bayesian optimization on the \emph{High Intensity Proton Accelerator (HIPA)} and the \emph{Swiss Free Electron Laser (SwissFEL)}. Further, we address practically relevant challenges in the context of tuning particle accelerators, including step-size constraints and user feedback for hyper-parameter tuning.
\newline
Tuning of accelerators is a challenging task for several reasons. Typically, simulation models are not accurate enough to allow offline optimization, and the parameters have to be tuned empirically by evaluating different settings on the machine. Among many other considerations, the following points are of particular importance; including on the machines we used for experimentation.%
\begin{description}
	\item[\normalfont \textbf{Safety}] Beam optimization is a delicate task because improper settings can cause beam losses that potentially damage the machine. While modern machines have safety measures implemented that are designed to prevent hardware damage by stopping the beam, it is desirable to avoid triggering such safety measures. On one hand, re-starting the machine is time-consuming and breaching the safety level is often considered off-limit. Further, an effective method should concentrate on sampling feasible points for fast convergence.
	\item[\normalfont \textbf{Step-Size Control}] Classical optimization methods are often based on the principle of making small adjustments to an incumbent solution, such as a small step in the gradient direction. Global search methods are an attractive alternative since they do not require gradient estimates and make less restrictive assumptions such as convexity of the target functions. Popular choices include Nelder-Mead \cite{nelder1965simplex}, CMA-ES \cite{hansen2003reducing} and Particle-Swarm optimization \cite{kennedy1995particle}.  However, these methods do \emph{not} impose a step-size constraint. Large changes to input parameters of particle accelerators can be problematic because oftentimes feedback systems are in place to stabilize the beam, which might not be able to follow abrupt changes. This limitation can be circumvented by slowly changing the machine parameters to the requested target values. Such a process, however, increases the measurement time on the machine and slows down the optimization.
	\item[\normalfont \textbf{User Feedback}] Optimization methods are usually not designed to provide user feedback other than the progress on the target value. In particular, it is difficult to monitor the optimization trajectory and the user essentially has to trust the method to not induce an unwanted state on the machine. 
\end{description}

As our main contribution and addressing the challenges outlined above, we demonstrate feasibility and efficacy of {\em safe Bayesian optimization} for parameter tuning on particle accelerators. Our approach is based on the \textsc{LineBO} method of \citet{kirschner2019linebo}, and we develop a version of the same method with step-size control. The \LineBO{} approach combines Bayesian optimization with a line-search technique, thereby allowing to scale the method to high-dimensional settings. We compare performance with the non-safe global search methods, CMA-ES \cite{hansen2003reducing} and Nelder-Mead \cite{nelder1965simplex}. Our claims are corroborated with experimental data on two machines: The High Intensity Proton Accelerator (HIPA, introduced in section~\ref{sec:HIPA}) and the Swiss Free Electron Laser (SwissFEL, introduced in section~\ref{sec:SwissFEL}). 

\subsection{A Brief Overview on Optimization Methods}

For many accelerators, parameter tuning has been automated to achieve significant speedups compared to manual operator tuning~\cite{agapov2014ocelot,tomin2016progress,tomin2017line}.
Since only point evaluations of the objective function are available, one has to
rely on so-called {\em zero-order} or {\em black-box} optimization methods. 
There are a large variety of optimization methods in use.
Due to their simplicity, the Nelder-Mead (Simplex)~\cite{nelder1965simplex} algorithm
and random walk optimizers have become popular choices
to assist operator tuning~\cite{huang2018, tomin2016progress, aiba2012}. 
Further methods that have been successfully applied on accelerators include Extremum Seeking~\cite{scheinker2013ES, scheinker2018ES, scheinker2019ES} and robust conjugate direction search (RCDS)~\cite{huang2013RCDS, huang2015RCDS}. 
The advantage of local descent methods is their simplicity, although convergence guarantees only apply under strong assumptions like convexity. Other gradient-based optimization algorithms are more difficult to deploy as additional samples are required to estimate the gradient, e.g.\ via finite differences. 

Another class of zero-order optimization methods are evolutionary algorithms such as Particle-Swarm optimization~\cite{huang2015RCDS} and genetic algorithms~\cite{tian2014, huang2014, bergan2019}. A popular choice due to its simplicity is CMA-ES \cite{hansen2003reducing}, which is shown to achieve competitive performance on standardized benchmarks \cite[e.g.][]{liu2020versatile}, although we are not aware of an application on particle accelerators.

As an alternative, {\em Bayesian optimization}~\cite{shahriari2015taking, frazier2018tutorial} has recently gained
interest in the accelerator community~\cite{kirschner2019swissfel,duris2020bayesian,hanuka2021PRAB,roussel2021PRAB}.
Bayesian optimization is a framework for zero-order optimization that deals with observational noise in a principled
way, allows the use of prior data or knowledge about the objective, and comes with theoretical convergence guarantees for some variants~\cite{shahriari2015taking}.

When tuning particle accelerators, there are often safety-relevant  constraints, e.g., physical or safety limits that cannot be violated, of which the typical optimization algorithm is unaware.
For example, at the Swiss Free Electron Laser the pulse energy should be maximized while keeping low undulator losses. At the High Intensity Proton Accelerator the situation is more rigid: It is important to keep the overall losses as low as possible while not exceeding any of the individual loss limits; violating the limit of a single loss monitor triggers the safety system, causes downtime, or possible damages to the machine.

It is natural to formulate this setup as a \emph{constrained optimization} problem, where the goal is to optimize a target function subject to a feasibility condition.  Established algorithms for the constrained setting include Sequential Quadratic Programming \cite[SQP][]{boggs1995sequential} and interior point methods \cite{byrd1999interior}. Variants for constrained Bayesian optimization have been proposed in \cite[e.g.][]{gardner2014bayesian,hernandez2016general,eriksson2021scalable}. Gradient descent methods can be applied in the constrained setting via a  Lagrangian (primal-dual) formulation of the objective that penalize infeasible points \cite{wei2020online}. We emphasize that the methods for constrained optimization ought to return a feasible \emph{final} solution but evaluating \emph{infeasible} parameters is tolerated during optimization; hence these methods do \emph{not} address the safety requirement for tuning particle accelerators. This is opposed to \emph{safe optimization}, where the complete optimization trajectory has to satisfy the feasibility condition. To achieve \emph{safe} optimization that guarantees feasibility for all evaluation points with high probability, \citet{Sui2015} proposed \emph{Safe Bayesian Optimization}. We describe this method and extensions in detail in section~\ref{sec:BayOpt}. A variant of Safe Bayesian Optimization was evaluated on SwissFEL in previous work~\cite{kirschner2019linebo}.

\subsection{HIPA}\label{sec:HIPA} 

The high intensity proton accelerator (HIPA) provides the primary beams to PSI's versatile experimental facilities.
HIPA generates a
proton beam with 590~MeV kinetic energy and presently
1.3~MW average beam power~\cite{Seidel2010}. 
This corresponds to a proton current of about 2.2~mA.
The HIPA accelerator consists of a Cockcroft-Walton
pre-accelerator and a chain of two isochronous
cyclotrons, the Injector II and the Ring cyclotron. The
beam is produced in continuous wave (CW) mode at a
frequency of 50.6~MHz. The high
intensity proton beam is used to produce pions and muons
by interaction with two graphite targets. 
After collimation the
remaining beam with roughly 1~MW is then used to
produce neutrons in a spallation target.
A pulsed source for ultracold neutrons (UCN) is also in operation~\cite{UCN2009}.

In practice, the
performance is limited by the beam losses at the
extraction of the Ring cyclotron and in the high energy beamline downstream. 
In the Ring cyclotron longitudinal space charge leads to an increase
in energy spread that transforms into transverse beam
tails, which reduces the turn separation at extraction~\cite{Seidel2010}.
Furthermore, significant emittance growth occurs
in the graphite targets and results in unavoidable
collimation loss in specially shielded collimators.
These losses are controlled by several RF and orbit feedback systems to keep the machine stable and by empirical operator tuning.


\subsection{SwissFEL}\label{sec:SwissFEL} 
The Swiss free electron laser (SwissFEL) is PSI`s state of the art source for ultra-short X-ray pulses~\cite{SF1, SF2}. A photocathode is used to generate short electron pulses which are then brought up to 6~GeV by a series of linear accelerator modules and compressed down to a pulse length of only a few femtoseconds in two bunch compressor chicanes. Inside an undulator line, the electron pulses self-modulate and emit X-ray pulses with laser-like properties by a process called self-amplified spontaneous emission (SASE). These ultra-short X-ray pulses are then used for various user experiments (e.g. pump probe experiments, X-ray spectroscopy). At SwissFEL, there are two undulator lines - Aramis and Athos - that can be used to generate photons for different end stations where the user experiments take place.

The settings of the accelerator (e.g. the photon energy or the pulse compression) are often adjusted to fit requests of the users. Typically subsequent parameter optimization is necessary in order to maximize the photon pulse energy at the experiment. The pulse energy is measured with a gas detector~\cite{SFGAS} that provides readings on a shot-to-shot basis. When performing the optimization, various tuning parameters ($\sim{100}$) can have a strong impact on the resulting beam quality, but also on the losses that can cause radiation damage to delicate equipment. Especially, in the undulator region, losses should be kept low while tuning on the SwissFEL output, to avoid long-term damage due to demagnetization of the permanent magnets inside the undulators.


\section{Bayesian Optimization with Constraints}\label{sec:BayOpt}

Bayesian optimization (BO) \citep{Mockus1982,shahriari2015taking,frazier2018tutorial} is a flexible, data-driven approach for global optimization with noisy feedback. 
The main idea is to fit a statistical regression model on the target function, using the data collected during optimization, and possibly, any available prior data. The next evaluation point is then chosen to reduce the model uncertainty about the unknown optimal solution. The approach was shown highly effective in many real-world applications, including optimization of particle accelerators \cite{kirschner2019linebo,duris2020bayesian,hanuka2020physics}. Variants for safety-critical tuning were developed by \cite{Sui2015,Duivenvoorden2017SafeOptSwarm,Berkenkamp2016BayesianSafety}, where the algorithm learns a model of additional constraints online. The constraint models are used to estimate a feasible `safe' set of parameters. The method then only evaluates such safe inputs and thereby ensures that no query point violates the safety constraints (with high probability). 

\subsection{Line Bayesian Optimization}

\begin{figure}
	\includegraphics{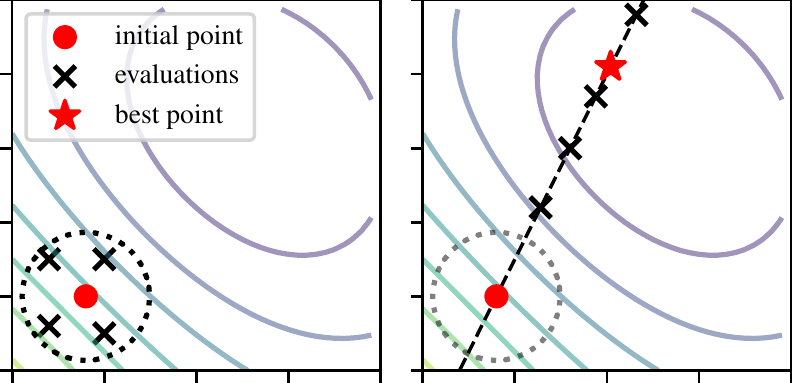}
	\caption{The \ALineBO~algorithm alternates between a local search (left; line \ref{ll:grad-start} to \ref{ll:grad-stop} in Algorithm \ref{alg:bayesopt}) and a line search (right; line \ref{ll:line-start} to \ref{ll:line-stop} in Algorithm \ref{alg:bayesopt}). On the left the search domain is restricted to a ball centered around the initial point. On the right, the search domain is chosen as a line in the direction of the estimated steepest ascent. The procedure is repeated at the best point found during the line search. }
	\label{fig:linebo}
\end{figure}

\LinesNumbered
\RestyleAlgo{ruled}
\newcommand{\REG}{\texttt{REGRESSION}}
\newcommand{\MINT}{\texttt{MACHINE-EVAL}}
\newcommand{\PLOT}{\texttt{LINE-PLOT}}
\newcommand{\ACQ}{\texttt{SAFE-ACQUISITION}}
\newcommand{\CANDIDATE}{\texttt{CANDIDATE}}
\newcommand{\MEAN}{\normalfont\texttt{MEAN}}
\newcommand{\UCB}{\normalfont \texttt{UCB}}
\newcommand{\LCB}{\normalfont\texttt{LCB}}
\newcommand{\CW}{\normalfont\texttt{CW}}
\newcommand{\model}{\hat m}
\newcommand{\eval}[2]{{\text{#1}_{#2}}}
\let\oldnl\nl
\newcommand{\nonl}{\renewcommand{\nl}{\let\nl\oldnl}}

In the following, we introduce a variant of safe Bayesian optimization developed in \cite{kirschner2019linebo}. The main idea is to alternate between two phases (see Figure~\ref{fig:linebo}). First, data is collected within a small ball around the current incumbent solution. Then a line search is performed in the direction of the (predicted) largest increase. In both phases, the data points are chosen to maximize an \emph{upper-confidence bound (UCB)} of the regression estimate, which is known to be an effective search surrogate \cite{Srinivas2009}. This procedure is repeated until a satisfactory solution is found, or the search budget is exhausted.
The intuition for these choices is twofold: First, by concentrating samples on a 1-dimensional domain or a small enough ball we obtain reliable function estimates with few samples in a local region of interest. Second, determining the acquisition points is computationally much simpler compared to standard Bayesian optimization, which requires to solve a non-convex optimization problem over the full input domain in each round.

\begin{figure*}
	\includegraphics[width=\textwidth]{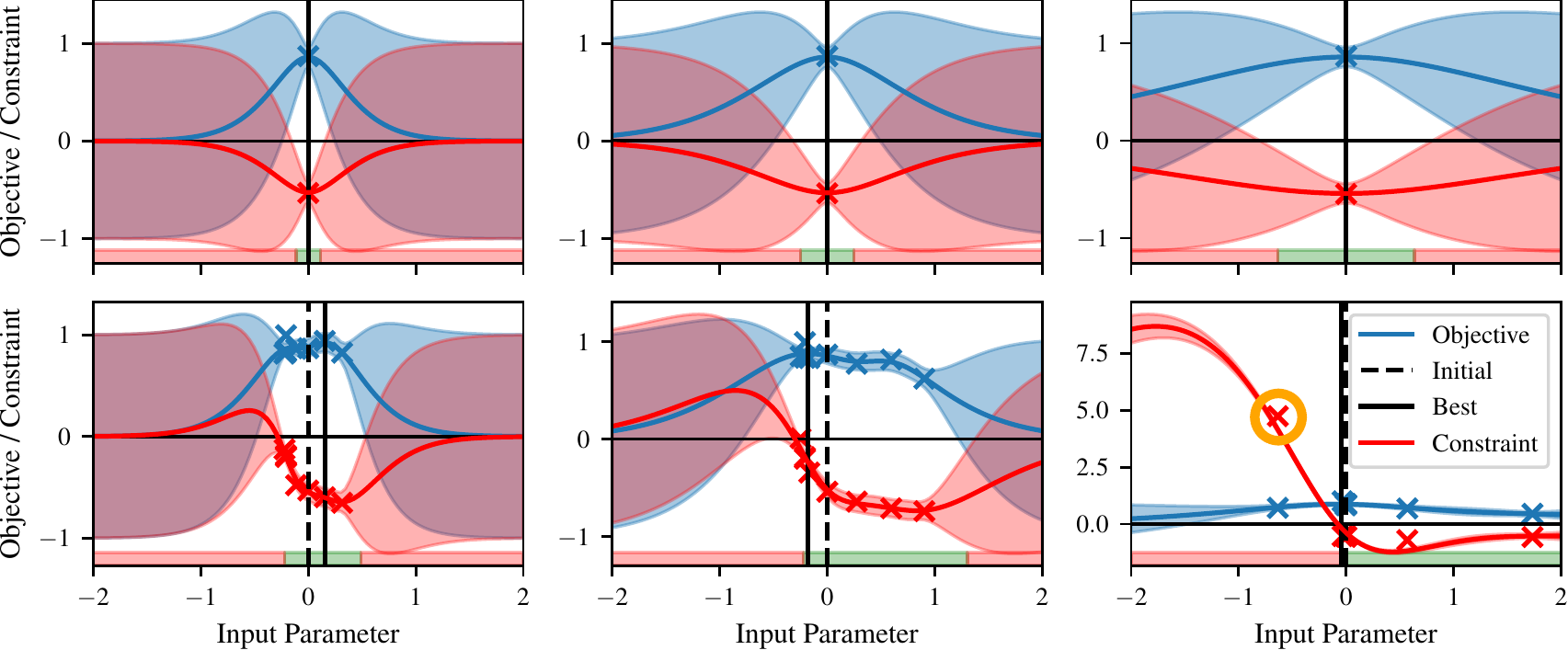}
	\caption{
		%
		%
		%
		%
		The plots show 1-dimensional slices of a synthetic objective (blue, maximized) and constraints (red). The left, middle and right column correspond to different GP models with lengthscales $0.1$, $0.2$ and $0.5$, respectively, on the same objective. We produce such plots live during optimization to evaluate model fit and confidence bands relative to the measurements (crosses). The vertical dashed line indicates the best parameter prior to optimizing the 1-dimensional subspace, and the vertical solid line indicates the updated best parameter. Green and red intervals on the x-axis correspond to predicted safe and unsafe parameters. Note that the safe region is expanded as more data is collected (cf.~top to bottom plot). When choosing hyperparameters such as the kernel lengthscale, system experts can visually inspect the predicted safe set, and make adjustments accordingly. A smaller lengthscale leads to a more conservative method that expands the safe set in smaller steps. Note that in the right column, the lengthscale is chosen too large. At the first evaluation the safeset is predicted incorrectly (see upper right plot), which consequently leads to an unsafe query point (orange circle, bottom plot). 
	}
	\label{fig:alg}
\end{figure*}

To formally introduce the algorithm, we write the tuning objective (maximized) and $l$ constraints as follows:
\begin{align*}
	\max_{x \in \xX} f(x) \quad \text{s.t.} \quad  \begin{array}{cc}
		g_1(x) \leq& 0 \\
		\vdots& \\
		g_l(x) \leq& 0 \\
	\end{array}
\end{align*}
Here, $f(x)$ is the unknown target function with input domain $\xX \subset \RR^d$, e.g.\ $\xX = \{ x \in \RR^d  :  0 \leq x_i \leq 1 \}$, corresponding to the set of $d$ tuning parameters on an allowed range. The functions $g_1(x), \dots, g_l(x)$ are constraints that need to be satisfied during operation, such as a loss limit. Note that the $g_1(x),\dots, g_l(x)$ are assumed to be unknown initially (like the target $f(x)$), and the only way of accessing the constraint functions is by evaluating a parameter on the machine. 
Provided an input parameter $x_t \in \xX$ at step $t$, the machine interface (denoted $\MINT(x_t)$) returns a set of noisy measurements,
\begin{align*}
	y_t &= f(x_t) + \epsilon_t\\
	z_{t,1} &= g_1(x_t) + \xi_{t,1}\\
	&\vdots\\
	z_{t,l} &= g_l(x_t) + \xi_{t,l}
\end{align*}
where $\epsilon_t$ and $\xi_{t,1}, \dots, \xi_{t,l}$ is measurement noise.
In practice, we average multiple evaluations to reduce the noise variance. Also note that the constraint limit is zero without loss of generality, because we can always redefine a constraint $g_i(x) \leq c$ as $\tilde g_i(x) =g_i(x) - c \leq 0$. We say that an evaluation procedure is \emph{safe} if it guarantees that with very high probability, all evaluation points satisfy $g_i(x_t) \leq 0$ for all $i=1,\dots,l$ and time steps $t \geq 1$.

The next step is to compute a regression model given the available data $\dD_t = \{(x_s, y_s, z_{s,1}, \dots, z_{s,l})\}_{s=1}^t$. In general, this is a user choice which we abstractly encapsulate in the $\REG(\cdot)$ oracle function. Given the data $\dD_t$, the regression oracle produces a model estimate $\hat m_t$. We assume that the model provides the following estimates for the target function and the constraints ($h \in \{f, g_1, \dots, g_l\}$):
\begin{itemize}
	\item $\eval{\MEAN}{h}(\model_t, x)$: The mean estimate of the target function evaluated at $x$; typically a consistent estimator of $h(x)$.
	\item $\eval{\UCB}{h}(\model_t, x, \delta)$: An upper confidence bound (UCB) on the true function, which takes a confidence parameter $\delta \in [0,1]$ and satisfies:
	\[\PP[\forall x \in \xX : h(x) \leq \eval{\UCB}{h}(\model_t, x, \delta)] \geq 1 - \delta\,,\]
	where the randomness is due to noisy evaluations. 
	\item $\eval{\LCB}{h}(\model_t, x, \delta)$, a lower confidence bound (LCB) on the true function, defined analogously to $\eval{\UCB}{h}$.
	\item $\eval{\CW}{h}(\hat m_t, x, \delta) \eqdef \eval{\UCB}{h}(\model, x, \delta) - \eval{\LCB}{h}(\model, x, \delta)$, the width of the confidence band at $x \in \xX$.
\end{itemize}
The most common regression model used in Bayesian optimization is Gaussian process (GP) regression (or kernel least-squares) \cite{Rassmussen2004}, which we also use in our experiments. A brief, illustrative introduction to GP-regression and the exact specification of our regression function is given in Appendix~\ref{app:gp}.

Finally, Bayesian optimization acquires points that are targeted to reduce the model uncertainty of the true maximizer $x^* = \argmax_{x \in \xX} f(x)$. To account for the safety constraints, at any step $t$, the acquisition is restricted to the \emph{safe set} with margin $\tau \geq 0$,
\begin{align*}
	\sS_t^\tau = \{ x \in \xX : \max_{i \in [l]} \eval{\UCB}{g_{i}}(\hat m_t, x, \delta) \leq - \tau\}\,.
\end{align*}
The safe set is defined as the subset of the domain $\xX$ where with probability at least $1-\delta$, the model does not predict a constraint violation with margin $\tau$. A larger safety margin makes the algorithm more conservative. Also note that the model predicts the average constraint level for a given input parameter. Therefore a positive safety margin is needed when the constraint measurements are noisy and the goal is to avoid constraint violations with high probability (i.e., controlling the tail of the constraint distribution).

The candidate solution $\hat x^*_i$ in iteration $i$ is defined as the maximizer of the mean estimate on $\sS_t^\tau$,
\begin{align*}
	\hat x^*_i = \CANDIDATE(\sS_t^\tau, \hat m_t) \eqdef \argmax_{x \in \sS_t^\tau} \eval{\MEAN}{f}(\hat m_t, x)
\end{align*}
Evaluating \emph{just} the incumbent solution however does not induce enough exploration, and the algorithm would get stuck in local optima. Instead, the next sequential measurement is chosen to maximize the upper confidence bound score function, which is a well-behaved surrogate to reduce the uncertainty of the true maximizer:
\begin{align}
	x_t = \argmax_{x \in \sS_t^\tau} \eval{\UCB}{f}(\hat m_t, x, \delta)\label{eq:ucb}
\end{align}
This choice leads to evaluation of points where the model predicts either a high reward or large uncertainty (or both). Upper confidence bound algorithms are widely studied in the literature \cite{auer2002confidencebounds} and convergences guarantees for these algorithms are known for the unconstrained case \cite{Srinivas2009,Chowdhury2017}. The constrained case requires additional care to ensure that model obtains sufficient data to expand the safe set \cite{Sui2015}. The exact acquisition procedure that we use is specified in Algorithm~\ref{alg:acq}. Since finding a maximizer of the acquisition score \eqref{eq:ucb} is challenging in high-dimensional domains, we rely on the \LineBO{} approach by \citet{kirschner2019linebo}. Instead of searching the full domain, \LineBO{} restricts the search space to an adaptively chosen sequence of one-dimensional subspaces. Let $\aA(x, w) = \{ x + a w : a \in \RR \}$ be an affine subspace for offset $x \in \RR^d$ and direction $w \in \RR^d$. To determine direction and offset we first collect data points in a small ball $\bB(\hat x^*_{i-1}, \eta)$ around the previous incumbent solution $\hat x^*_{i-1}$ to compute a new candidate $\hat x^*_i$. The offset is then set to the new candidate $\hat x^*_i$ and the search direction is chosen as $w = \hat x^*_i - \hat x^*_{i-1}$.

For step-size control during the line-search phase, the algorithm constrains the next evaluation point $x$ at any time to $\bB(\hat x^*_i, \eta)$. The complete approach is summarized in Algorithm \ref{alg:bayesopt}. An illustration is given in Figure~\ref{fig:linebo} and the line search phase is shown in Figure~\ref{fig:alg}.

\begin{algorithm2e}[ht]
	\DontPrintSemicolon
	\SetAlgoVlined
	\SetAlgoNoLine
	\SetAlgoNoEnd
	
	\KwIn{domain $\xX$, initial evaluation point $\hat x_0$, initial data $\dD_0$, confidence $\delta \in [0,1]$, step-size $\eta > 0$, safety margin $\tau \geq 0$, $j_{ball}, j_{line} \in \NN$ inner loop evaluations
	}
	\KwOut{Estimated maximizer}
	\caption{\ALineBO } \label{alg:bayesopt}
	
	$t \gets 1,\quad i \gets 0,\quad \hat x^*_0 \gets x_0$ \tcp*{initialization}
	\While{not stopped by user}{
		\tcp{Determine best search direction $w_i$}
		\label{ll:grad-start}\For{$j = 1,\dots, j_{ball}$}{
			$\model_{t} \gets \REG(\dD_{t-1})$\;
			$x_t \gets \ACQ(\bB(\hat x^*_{i-1}, \eta) \cap \xX, \model_t, \tau)$\;
			$\dD_t \gets \dD_{t-1} \cup (x_t, \MINT(x_t))$\;
			$t \gets t+1$\tcp*{evaluation counter}
		}
		\tcp{Update candidate and search direction}
		$\hat x_i^* \gets \CANDIDATE(\bB(\hat x^*_{i-1}, \eta) \cap \xX, \model_t, \tau)$\;
		\label{ll:grad-stop}	$w_i \gets \hat x_i^* - \hat x_{i-1}^*$\;
		\tcp{Search line in direction $w_i$ \& offset $\hat x^*_i$}
	\label{ll:line-start}	\For{$j = 1,\dots, j_{line}$}{
			$\model_{t} \gets \REG(\dD_{t-1})$\;
			$\hat x_i^* \gets \CANDIDATE(\aA(\hat x^*_i, w_i) \cap \bB(\hat x_i^*, \eta)\cap \xX, \model_{t}, \tau)$\;
			\label{ll:acq-line}$x_t \gets \ACQ(\aA(\hat x^*_i, w_i) \cap \bB(\hat x_i^*, \eta) \cap \xX, \model_t, \tau)$\;
			$\dD_t \gets \dD_{t-1} \cup (x_t, \MINT(x_t))$\; 
			\tcp{Optional user feedback:}
			$\PLOT(\dD_t, \model_t, \hat x_t, w_i)$\;
			$t \gets t+1$\tcp*{evaluation counter} \label{ll:line-stop}
		}
		$i \gets i+1$\tcp*{next iteration}
		
	}
	\Return{$\hat x^*_{i-1}$}
\end{algorithm2e}

\begin{algorithm2e}[ht]
	\DontPrintSemicolon
	\SetAlgoVlined
	\SetAlgoNoLine
	\SetAlgoNoEnd
	\newcommand{\xsearch}{\tilde \xX}
	\newcommand{\xexpand}{x^{\text{\normalfont EXP}}}
	\newcommand{\xucb}{x^{\text{\normalfont UCB}}}
	\newcommand{\xucbsafe}{x^{\text{\normalfont UCB-S}}}
	
	\KwIn{search domain $\xsearch$, model $\model$, confidence $\delta$, safety margin $\tau$}
	\KwOut{acquisition point $x$}
	\caption{\ACQ} \label{alg:acq}
	\tcp{Define safe-set:}
	$\sS^\tau \gets \{ x \in \xsearch  : \max_{i \in [l]} \eval{\UCB}{h_{i},\delta}(\model, x) \leq -\tau \}$\;
	\tcp{UCB and Safe UCB parameter:}
	$\xucb \gets \argmax_{x \in \xsearch} \eval{\UCB}{f}(\model, x, \delta)$\;
	$\xucbsafe \gets \argmax_{x \in \sS^\tau} \eval{\UCB}{f}(\model, x, \delta)$\;
	\If{$\xucb = \xucbsafe$}{
		\Return{$\xucbsafe$}
	}
	\tcp{Compute expanding point towards $\xucb$:}	
	$\xexpand \gets \argmin_{x \in \sS} \|x - x^{\text{UCB}}\|$\;
	\If{$\max_{i \in [l]} \eval{\CW}{g_i}(\model, \xexpand, \delta) >\eval{\CW}{f}(\model, \xucbsafe, \delta)$}{
		\Return{$\xexpand$}
	}
	\Else{
		\Return{$\xucbsafe$}
	}
\end{algorithm2e}

\subsection{Variants}

In our experiments, we evaluate the following variants of the proposed approach:
\begin{description}
	\item[\normalfont \ALineBOloc] This is the main (ascent \& localized) variant shown in Algorithm \ref{alg:bayesopt}. We use $\eta = 0.1$ relative to a domain that is normalized to $[0,1]^d$. 
	\item[\normalfont \ALineBO] For this variant, we do not enforce the step size constraint in Line \ref{ll:acq-line} of Algorithm~\ref{alg:bayesopt}, which corresponds to the variant proposed in \cite{kirschner2019linebo,kirschner2019swissfel}.
	\item[\normalfont \CLineBOloc] This variant uses coordinate aligned search directions, also known as \emph{coordinate descent}. Specifically, we replace Line \ref{ll:grad-start}-\ref{ll:grad-stop} in Algorithm \ref{alg:bayesopt} by setting $w_i = e_{i \bmod d}$, where $e_1,\dots,e_d$ are the basis vectors.
	\item[\normalfont \CLineBO] Similar to \ALineBO{}, this variant does not enforce the step size constraint in Line \ref{ll:acq-line}. Note, this variant (without the step-size constraint) was proposed and evaluated by \citet{kirschner2019linebo,kirschner2019swissfel}.
\end{description}

\subsection{Choice of Hyperparameters}

Algorithm~\ref{alg:bayesopt} relies on several hyperparameters that are configured by the user. We first explain our choices in the following. Then we briefly discuss more general methods for adjusting the hyperparameters parameters.

For the number of evaluations per round, we set $j_{ball} = 2d$ and $j_{line} = 10$. From our experience, this allows to estimate a reasonable search direction and make sufficient progress on each line (and the method can revisit the same search direction in the next iteration). When the signal is very noisy, a larger number of evaluations per iteration can be appropriate. On systems with a large number of hyperparameters ($d > 50$) but low \emph{effective dimensionality}, it can be effective to choose search direction uniformly at random \cite{kirschner2019linebo}. The step-size $\eta$ and margin $\tau$ are provided by the user. The exact choice is usually based on system considerations. We use a maximum step size of $0.1$ (i.e.,~$10\%$ of the allowed input range) and $\tau=0.1$ (i.e.,~$10 \%$ below the critical value).
 
Parameters of the GP model encode regularity assumptions on the objective and constraints. Choosing suitable model parameters is essential for the effectiveness of Bayesian optimization. In a first step, we normalize input ranges to obtain a normalized domain  $\xX = [0,1]^d$. Further, we normalize the observed objective values to $[0,1]$, and the range of feasible constraint values to $[-1,0]$. The latter is important to ensure that the prior upper-confidence bound predictions of the constraints are positive. This corresponds to assuming that an input parameter is unsafe in the absence of data. 

We then use a Matern52 kernel with lengthscale 0.2 and unit prior variance, which we found effective and robust choices across all our experiments. For the observation noise likelihood, we use a Gaussian distribution with an empirical variance estimate. We choose to keep all hyperparameters fixed during optimization, but adaptive choices are possible in principle.

The above choice of kernel and lengthscale is appropriate when a small change in the input parameters (e.g., $< 10\%$ relative to the input range) leads to a moderate change in the output, e.g., does not suddenly increase the constraint level to a critical value. This requires to choose suitable allowed ranges of tuning parameters, which we determined together with system experts. 

To adjust these parameters on a new system, the practitioner can inspect the model predictions after the initial (safe) point has been evaluated. Most importantly, the predicted safe set for each parameter can be plotted and evaluated by a system expert. Based on the visual feedback, the lengthscale or input ranges of each parameter can be adjusted to meet the operators expectation on the safety of the method. The method acts more conservatively for larger safety-margin $\tau$, smaller step-size $\eta$, and smaller lengthscale parameter used for the kernel estimation. The visual inspection procedure is illustrated in Figure \ref{fig:alg}.

We also produce the line plots online during the line-search phase. 
We found this to be an extremely valuable tool to further calibrate hyperparameters and obtain feedback on the optimization progress. 
Specifically, too conservative parameter choices (e.g., a very small lengthscale) lead to slow expansion of the safe set and no/little extrapolation of the target values beyond the data points. On the other hand, if the regression estimate appears too smooth relative to the observed data points, or the algorithm evaluates points close or above the constraint limit, more conservative hyperparameters are advisable. 

In applications where prior data or a physics simulation is available, an alternative approach is to estimate hyperparameters in advance, e.g., using a maximum likelihood or marginalized likelihood estimation \cite{Rassmussen2004}. For a successful approach to extract kernel hyperparameters from a physics model, see the work by \citet{hanuka2021PRAB}. Some previous work has also explored adaptive model selection, e.g. \cite{berkenkamp2019no}, although this is still largely an active research area today.

Lastly, a word of caution is unavoidable. Safety and effectiveness of the methods is subject to the modeling assumptions, and therefore, choice of hyperparameters. In the absence of any assumptions on the system, a principled choice is limited. The challenge is amplified for safe optimization, because no method can guarantee safety if the system behaves in a way not anticipated by the algorithm.

\begin{figure*}
	\includegraphics{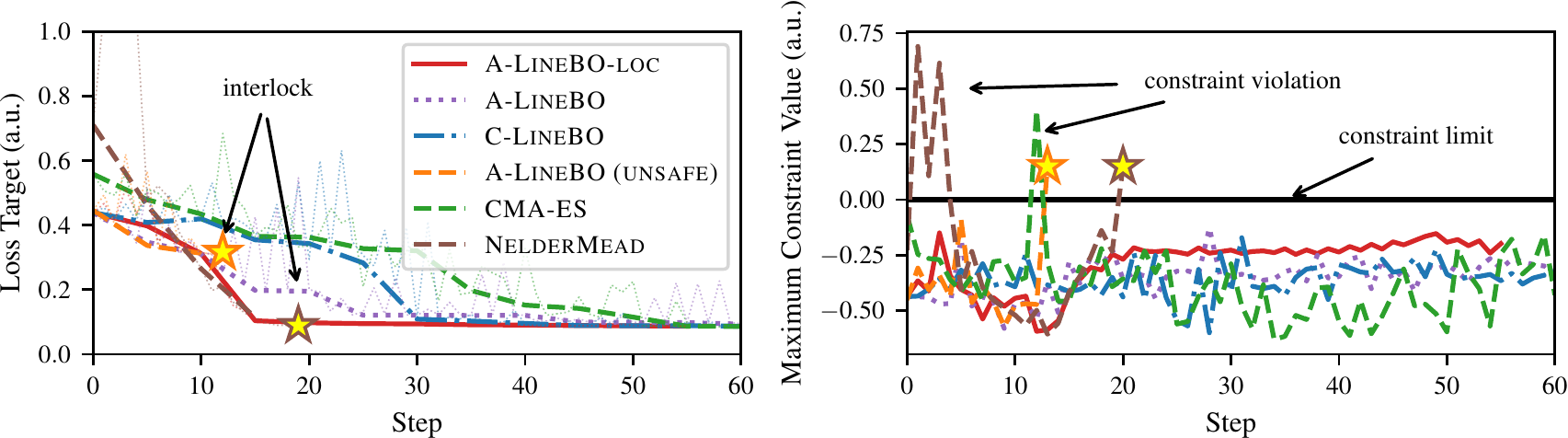}
	\caption{ Step-by-step evolution of the target function for the different optimization algorithms. We performed optimization at low beam intensity, using 5 quadrupoles strengths on a manually detuned initial point. The left plot shows the target values. The faint lines represent the measurement of the target function at each step. The solid lines depict the measurement of the target function at the best set of parameters predicted by the optimization models (updated only every 5 steps). The right plot shows the maximum constraint value at each step, where a value above 0 corresponds to a constraint violation. Note that \NelderMead{} and \ALineBOunsafe{} caused interlocks at step 12 and 19 respectively. Note that the exact values of the corresponding constraint violations are unknown, because the machine protection system stops beam operation before the measurement is registered. \CMAES{} and \NelderMead{} caused constraint violations that did not trigger interlocks at steps 12 and 2-4 respectively.}
	\label{fig:alg_comparison}
\end{figure*}

\subsection{Remarks on Computation}\label{ss:computation}

GP regression requires to compute the inverse of the kernel matrix, which can be done in $\oO(t^3)$ basic operations per round. With incremental updates, the complexity is reduced to $\oO(t^2)$. In practice, this effectively limits the number of data points that can be handled without significantly increasing computation time to $\sim 1000\text{ - }2000$. See Appendix \ref{app:gp} for further comments.

The \ACQ{} subroutine requires solving non-convex optimization problems over continuous domains. When the domain is one-dimensional as in the line search, a fine uniform (equidistant) discretization of the domain easily produces a solution below the system accuracy / noise level, such that the approximation error translates to a negligible difference on the objective value (we used 300 points per line). Maximizing the acquisition score within in the ball $\bB(\hat x^*, \eta)$ is computationally expensive in general. Nevertheless, since the domain is relatively small compared to the full domain, we found random search to be effective in our experiments. We compute the UCB score at 500 points sampled uniformly from $\bB(\hat x^*, \eta)$ and choose the maximizer among these. An alternative is to maximize the UCB score using any standard optimization algorithm such as gradient descent with random restarts.

Lastly, we note that in time-critical applications with a large number of constraints, a significant speedup can be obtained by using a joined covariance matrix for all constraints. Further, the most expensive step in computing the estimates is the inversion of the covariance matrix, which can be pre-computed while waiting for the machine to return the next evaluation.

\subsection{Limitations}
Enforcing a step-size sometimes limits the algorithm's ability to escape local optima. However, theoretical analysis suggests that such an approach at least finds a local optimum \cite{combes2014unimodal,saber2020forced}. In our experiments we did not see any performance degradation from enforcing a step-size (see Section \ref{sec:Experiment}). This indicates that, perhaps surprisingly, high-quality local solutions exist in the optimization surfaces that we encountered. On systems where local optima prevent the algorithm from progressing, one can use multiple starting points and drop the step-size constraint during the line-search phase. Another option is to replace the line search by a search over higher-dimensional subspaces. \looseness=-1

Another challenge is model misspecification. If a parameter is incorrectly predicted as safe, the parameter can become infeasible under the model's predictions \emph{after} the evaluation (c.f.~Figure~\ref{fig:alg}, right column). With a small step-size, this can lead to a situation where the predicted safe set is empty. A simple resolution is to back-track and restart the method from an earlier safe point. In such cases, it is further advisable to choose a smaller lengthscale, to prevent evaluation of infeasible points in the first place. 
 
\section{Experimental Evaluation}\label{sec:Experiment}

We first present empirical results on HIPA, and then an additional experiment on SwissFEL below.

\begin{figure}[htbp]
	\includegraphics{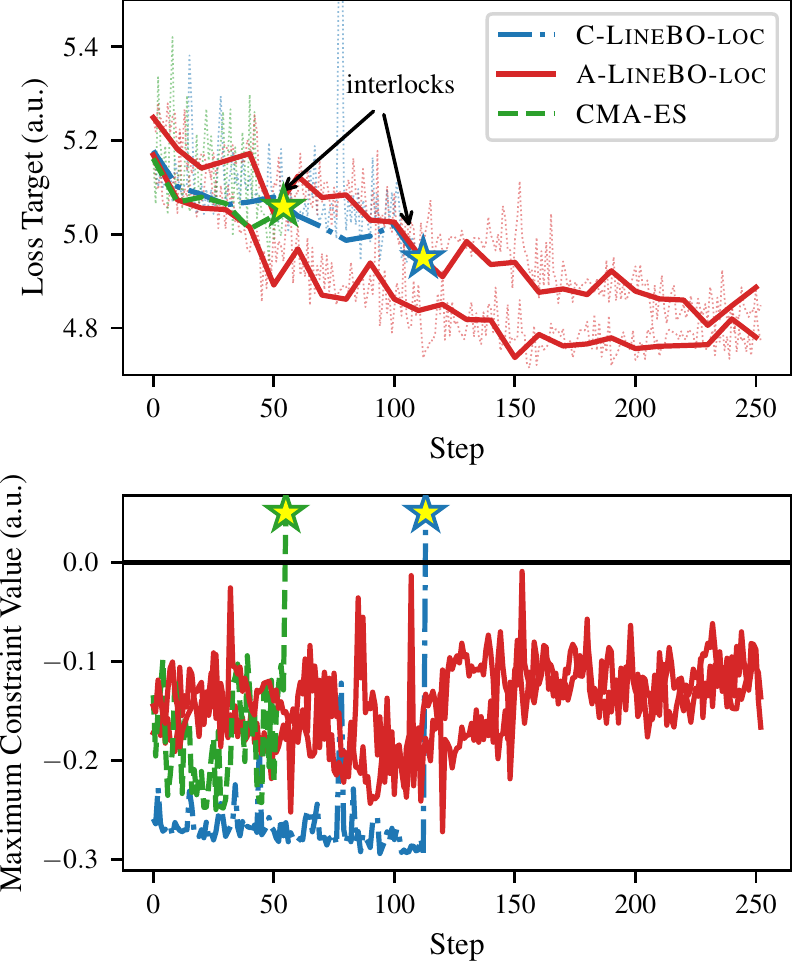}
	\caption{(HIPA)~~Experiment at high beam intensity with 16 tuning parameters. Both \CMAES\ and \CLineBOloc\ caused interlocks, and the run was stopped after 50 and 110 steps respectively. Note that the exact constraint value of the constraint violation is not known as the machine protection system prevents the measurement.
	\ALineBOloc\ successfully tunes the machine at high intensity (repeated twice). This experiment used an earlier configuration of the \LineBO~ methods, where we used a single constraint defined as the maximum over the 224 (normalized) loss signals and no safety margin ($\tau=0$). We think that this configuration led to misspecified confidence intervals, and therefore an interlock of $\CLineBOloc$. In such an event we recommend to use more conservative hyperparameters. }
	\label{fig:HIPA-full}
\end{figure}

\subsection{HIPA} \label{sec:exp_val_hipa}

The different versions of the BO algorithm with constraints have been tested in several dedicated experiments at HIPA.
The target of these experiments has been to reduce the overall beam losses around the machine by minimizing a target signal defined as a weighted sum of about 60 beam loss monitors (or loss related monitors) spread across the machine.
This weighted sum reflects that the beam loss monitors are not distributed equidistantly along the machine and that losses at lower energies cause less activation.

For each beam-related signal that could trigger a beam interruption (\textit{interlock}), a safety constraint function is defined as the difference between the signal and its beam interruption limit.
If available, the \textit{warning} limit of each signal is used and not the actual limit that would produce a beam interruption when crossed, which allows a safety margin in case the safety constraint are accidentally violated when operating near the limit.
For optimization of HIPA, we used 224 of such signals, mostly current and temperature measurements in collimators and beam loss monitors. We normalized each of the safety signals $g_i^{\text{loss}}(x)$ to define constraint functions
\begin{align*}
g_i(x) =  \frac{g_i^{\text{loss}}(x)- g_{i,limit}} {g_{i,limit}},
\end{align*}
where $g_{i,limit}$ is the limit of constraint $i$. Therefore the constraint reads $g_i(x) \leq 0$, and the safe outcome range is $[-1,0]$ (assuming that $g_i^{\text{loss}}(x) \geq 0$). For faster computation, it is possible to take the maximum over multiple constraints. However, this can lead to less smooth functions that are harder to learn by the GP model.

At HIPA, a beam current regulation and several beam orbit feedbacks are in continuous operation to ensure stable operation without drifts.
Every time a new evaluation point is set, the algorithms wait for these feedbacks to stabilize before acquiring data. Larger steps typically result in more time required until the machine is stable and the measurement is taken.

In our experimental validations, the variables used to optimize the target function are the field strengths of quadrupoles in the 870~keV transport beam line after the Cockcroft-Walton pre-accelerator towards the first of the two cyclotrons, the Injector II,
with an allowed empirical range of $\pm$2~A. This range is equivalent to an integrated field gradient of about $\pm$0.01~T, and is a typical range operators use for tuning.
In the center of Injector II the beam is collimated from 10~mA to about 2~mA with several collimators~\cite{Kolano2018}.
By adjusting the quadrupole strengths upstream, the beam can be shaped and matching in the cyclotron can be improved~\cite{Markovits87}, which will avoid beam tails and reduce beam losses throughout the accelerator downstream.


In order to protect the machine, the first experiments were performed at low intensity (\SI{100}{\micro\ampere}), which also allows a larger margin in the constraint functions limits as the losses are in general lower at low intensity. To give some headroom to the automatic algorithms and to simulate a sub-optimal machine, the selected quadrupole settings are manually moved away from the default settings. These detuned settings are recorded to have a consistent initial point.
In later sessions the algorithms were tested without detuning at full production current (about \SI{1930}{\micro\ampere}).

The left plot in Figure~\ref{fig:alg_comparison} shows the target evolution of the different algorithms at low beam intensity.
The faint lines show the evaluation of the target function at every new parameter configuration (step).
The solid line shows the measurement of the target function for the best set of parameters predicted by the algorithms after every 5 steps (evaluating the candidate solution more frequently slows down the overall optimization).
The right plot shows the constraint closest to the limit at each evaluation point (that is, the maximum over the normalized constraint levels). Note that NelderMead and \ALineBOunsafe{} caused beam interruptions due to constraint violations. These constraint violations are \emph{not} visible on the constraint plot because the machine projection system stops beam operation before the measurement is registered. Further, \CMAES{} and \NelderMead{} show constraint violations that did not lead to beam interruptions. This can happen in our setup when the warning level of a loss monitor is reached, but not the beam interruption level. All safe methods and \CMAES{} successfully optimize the loss to a final level of about 0.1, where \ALineBOloc{} converges the fastest.

\begin{figure*}
	\includegraphics{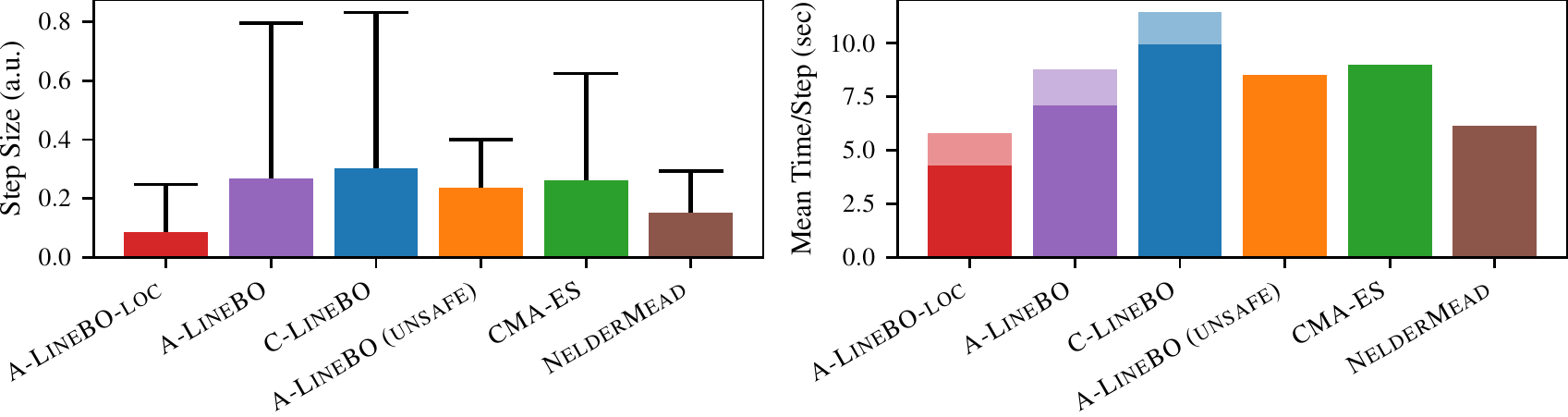}
	\caption{(HIPA)~~Observed average step-sizes (T-bars: maximum) for different algorithms measured in $l_2$-distance between parameter vectors with normalized input range (left). The right plot shows time used per step, where the solid part of each bar corresponds to machine evaluation time, and the faint part to compute time (on a single core Intel\textsuperscript{\copyright} Xeon\textsuperscript{\copyright} Gold 6140 2.3~GHz with 4~GB RAM), i.e.~\ALineBOloc: 1.54~s, \ALineBO: 1.72~s, \CLineBO: 1.51~s, \ALineBOunsafe:  0.01~s, \CMAES: 0.0004~s, \NelderMead:  0.0004~s. 
		The significant increase in computation time for the constraint method is due to the large number of constraints (224). This can be improved by using a joined covariance matrix for all constraints, see the remarks in \ref{ss:computation}. \ALineBO{} and \CLineBO{} are not step-size constrained, which leads to larger machine evaluation times. \CLineBOloc{} has an explicit step-size limit which leads to shorter machine evaluation times. Note that \ALineBOunsafe{} and \NelderMead{} caused interlocks after few steps (see Figure \ref{fig:alg_comparison}).}
	\label{fig:step_size_analysis}
\end{figure*}

Our second experiment at full production intensity is shown in Figure~\ref{fig:HIPA-full}, where we optimize 16 quadrupoles (all the quadrupoles in the \SI{870}{\kilo\electronvolt} beam line). The target reduction is less steep because, in order to avoid putting the machine in a dangerous state, in this case the parameters were not manually detuned and the initial point was already fairly close to the optimum. Notice also the difference on the magnitude of the target function due to the higher intensity and therefore higher losses around the machine.
The {\normalfont \CLineBOloc} and {\normalfont \CMAES} algorithms stopped early as they triggered beam interruptions in their last parameter settings.

This experiment was performed with an earlier configuration of our method without safety margin ($\tau = 0$). Moreover, we used a single constraint defined as the maximum over the 224 loss monitors. Consequently, an interlock was caused by \CLineBOloc{}, either because of model-misspecification (e.g., the maximum over the constraints was non-smooth) and consequently invalid confidence intervals, or noise of a constraint monitor that was very close to its limit. In such cases we recommend more conservative hyper-parameters (bigger margin $\tau$, smaller lengthscales of the GP models), as was used in the experiment shown in Figure \ref{fig:alg_comparison}.
{\normalfont \ALineBOloc} completed the optimization successfully twice, without triggering any beam interruption.


Figure~\ref{fig:step_size_analysis} shows an analysis of the step sizes and their impact on the speed of the optimization of the different algorithms. The plots show data from the low intensity experiment. The {\normalfont \CLineBO} algorithm has no limitations on the step sizes and therefore is allowed to navigate the whole safe region in single steps.
Compared with the {\normalfont \ALineBOloc} algorithm with limits on its step size it can be seen on the left plot how the unrestricted algorithms takes much larger steps.
The impact of these large steps can be seen on the right column plot, as due to the large steps taken by the unrestricted algorithm the optimization took about $1.5 \times$ as long to complete because it takes the machine feedbacks and regulation systems much longer to converge for these large steps. We remark that the high computation time for the safe BO methods is due to the use of 224 constraints. We expect that significant speedups are possible when using a joint kernel matrix for all constraints. Further, the most expensive step of inverting the covariance matrix can be pre-computed while waiting for the machine to return the next measurement.

A closer analysis (see Appendix \ref{app:experiments}) shows that the step sizes taken by the {\normalfont \CMAES} diverge as the algorithm approaches the optimization solution, as the target function becomes close to invariant to the change of the parameters in this region and tries to spread the sampling points. This is an undesirable behavior with respect to potential constraint violations or unstable machine states caused by large steps.



In conclusion, the experiments at HIPA demonstrate that the BO optimization algorithms successfully reduce the losses around the machine in a safe and efficient way, both at low intensity and full intensity operational setting. 
Particularly, the {\normalfont \ALineBOloc} algorithm outperforms or matches {\normalfont \CMAES} and the loss level achieved by human operators, while staying in general below the limits that would trigger beam interruptions.
The constrained step sizes of this method also allow for faster and more stable tuning.


\subsection{SwissFEL} \label{sec:exp_val_swissfel}

Previous work \cite{kirschner2019swissfel} established that \LineBO{} outperforms \NelderMead{} \cite{nelder1965simplex} for tuning the beam intensity on SwissFEL. Here, we provide an additional experiment on SwissFEL, including \CMAES{}, {\normalfont \ALineBO}, and the step-size constrained variant {\normalfont \ALineBOloc}.
As mentioned in section~\ref{sec:SwissFEL} the tuning signal is a shot-by-shot signal of the gas detector that gives a signal proportional to the amount of photons in a single pulse.
As optimization variables, we use 10 (horizontal and vertical) beam position monitor
target values for the trajectory feedback in the undulator section. We used as constraints 16 individual loss monitors that were combined into a single constraint. During optimization with the specified parameters, we did not observe any losses.

\begin{figure}[tbp]
	\includegraphics{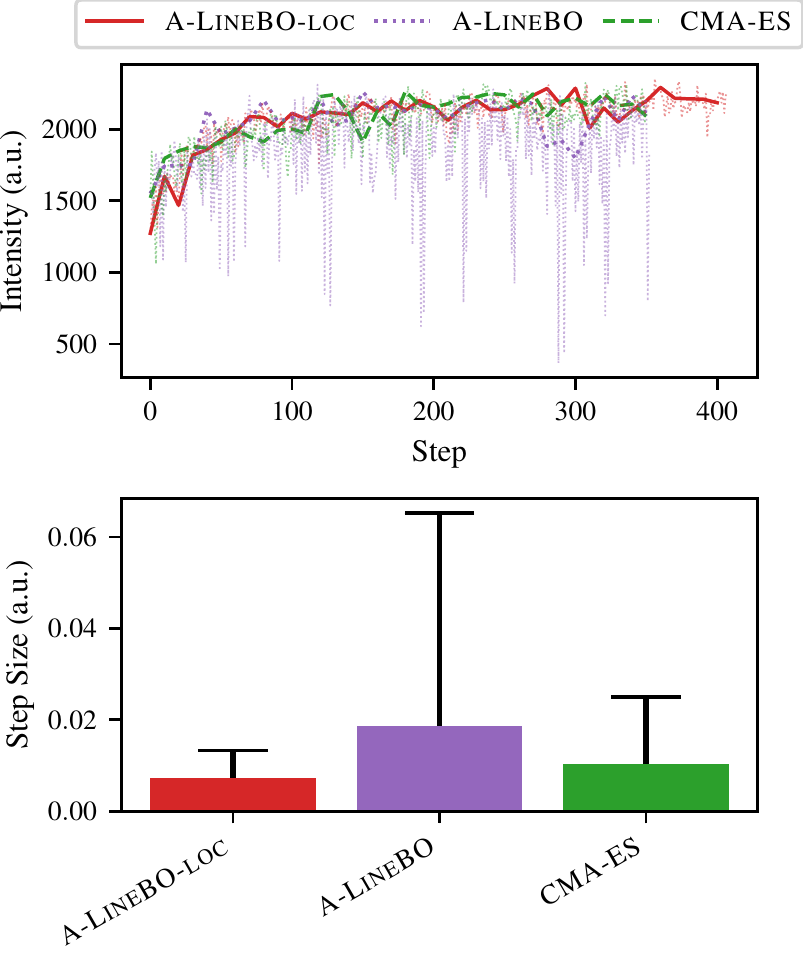}
	\caption{(SwissFEL)~~Step-by-step evolution of the SwissFEL single pulse intensity (top) and average step-size (T-bars: maximum) during optimization of 24 beam position monitor target values. The results demonstrate that the same methods can be used on different accelerators. We emphasize that, similar to the experiments on HIPA, the step-size constraints did not affect optimization performance while resulting in significantly smaller variation in the input space, and therefore a more stable machine. This can be seen in the top plot, where the faint lines depict the objective value observed at the evaluation points during optimization. }
	\label{fig:swissfel_analysis}
\end{figure}

The top plot of Figure~\ref{fig:swissfel_analysis} shows the target evolution of the different algorithms. 
The faint lines show the evaluation of the target function at every new parameter configuration (step).
The solid line shows again the measurement of target function for the best set of parameters predicted by the algorithms every 10 steps.
While all three tested algorithms achieve a similar final target value, it can be seen that the target value often drops for {\normalfont \ALineBO}. This is due to large steps in the parameter space, as shown in the bottom plot of Figure~\ref{fig:swissfel_analysis}, 
where the average (normalized) step-size is plotted.
It can also be seen that the {\normalfont \ALineBOloc} performs slightly better than {\normalfont \CMAES} in this respect.
Having a low variability and a high average of the target value  is especially important during {\it parasitic} tuning, i.e.\ tuning during user's experiments.
For this reason the step-size constrained {\normalfont \ALineBOloc} is the preferred algorithm.

\section{Conclusions}\label{sec:Conclusions}

While tuning particle accelerators remains a challenging task,
we demonstrated the feasibility of safe Bayesian optimization for automated parameter tuning. Safe Bayesian optimization does not rely on any specific machine model, and accounts for an arbitrary number of safety constraints.
By both optimizing and exploring the target and constraints functions, the algorithm continuously improves its estimate and uncertainty of these functions, and in this way uses all available data to evaluate promising but safe parameters. 
Multiple variants of safe Bayesian optimization have been described in detail,
including a newly developed variant with step-size control. This is an important feature for tuning particle accelerators,
since large parameter steps can lead to an unstable machine state as fast feedbacks might not be able to follow.
These variants have been applied and compared with a non-safe search methods CMA-ES, NelderMead and Bayesian optimization without constraints.
We presented experimental data for two accelerators, HIPA and SwissFEL, showing efficacy and reproducibility in several optimization runs.
Overall, we find that the different variants of Bayesian optimization and CMA-ES are both effective tools for beam optimization. 
However, safe Bayesian optimization is shown to cause constraint violations much less likely, and consequently, avoid beam interruptions.
In addition, the step-size controlled variants allow for a faster and more stable tuning, which is less disruptive for the users.

\begin{acknowledgments}
The authors would like to thank Rasmus Ischebeck for establishing the first contact, and his contributions to the earlier publications of the collaboration. Manuel Nonnenmacher and Andreas Adelmann contributed to early development and testing of the LineBO algorithm as part of Manuel's Master thesis. Nicolas Lehmann and Marco Boll contributed to the GUI development. We acknowledge the support of the SwissFEL and HIPA operation teams during the tuning experiments.

This research was supported by SNSF grant $200020\_159557$ and $407540\_167212$ through the NRP 75 Big Data program. Johannes Kirschner acknowledges funding through the SNSF Early Postdoc.Mobility fellowship P2EZP2\_199781. The project has received funding from the European Research Council (ERC) under the European Union’s Horizon 2020 research and innovation programme grant agreement No 815943.  This project was also supported by the Swiss Data Science Center (SDSC) within the Particle Accelerator and Machine Learning (PACMAN) project.
\end{acknowledgments}
\vfill

\newpage
\bibliography{references}

\appendix

\section{Gaussian Process Regression} \label{app:gp}

\paragraph{Kernel Regression} With $t$ evaluation-feedback pairs $\dD_t = \{(x_1, y_1), \dots, (x_t, y_t)\}$ available from previous steps or data collected prior to optimization, we can compute an estimate $\hat f_t$ to approximate the unknown true function, and the estimate is used to inform the next evaluation point. Most commonly, Bayesian optimization uses kernel least-squares regression to estimate the target and constraint functions:
\begin{align}
	\hat f_t = \argmin_{f \in \hH} \sum_{s=1}^t \big(f(x_s) - y_s)\big)^2 + \|f\|_{\hH_k}^2\,.
\end{align}
The minimization is over functions of a reproducing kernel Hilbert space (RKHS) $\hH_k$ \citep[see, e.g.~][]{Rassmussen2004}, which is defined by a kernel function $k : \xX\times \xX \rightarrow \RR$, and $\|\cdot\|_{\hH_k}$ denotes the associated Hilbert norm. The estimate $\hat f_t$ can be conveniently computed in analytic closed form as follows:
\begin{align}
	\hat f_t(x) = k_t(x)^\top (K_t + \mathbf{1}_t) ^{-1}y
\end{align}
where $\mathbf{1}_t \in \RR^{t\times t}$ is the identity matrix, $K_t$ is the $t\times t$ matrix containing $(K_t)_{ij}=k(x_i,x_j)$ formed by pointwise evaluation of the kernel function, $y$ is a vector of responses and lastly $k_t(x)_i = k(x,x_i)$ interpolates the measurement $x$ to other data points. A direct implementation requires the inversion of the kernel matrix $K_t$ with computational complexity $\oO(t^3)$. Iterative updates using the Sherman-Morrison identity reduce the complexity to $\oO(t^2)$ per round. If the computational burden is too large, methods relying on approximation tools for kernel spaces have been developed with near-linear dependence for low dimensional problems \cite{Mutny2018b,Calandriello2019}.

Frequentist confidence bands $\hat f_t(x) \pm \beta_{t,\delta}^{1/2} \sigma_t(x)$ have been derived in the literature, which contain the true function with high probability in the realizable case \cite{Srinivas2009,AbbasiYadkori2012}.
The quantity $\sigma_t$ corresponds to the estimation uncertainty and can likewise be derived in closed form, \begin{equation}
	\sigma_t(x) = \sqrt{k(x,x) - k_t(x)^\top K_t^{-1} k_t(x)}, 
\end{equation}
where $K_t$ and $k_t(x)$ as above. The parameter $\beta^{1/2}_{t,\delta}$ is a confidence parameter that determines the coverage and can be chosen according to frequentist theoretical results or tuned as a hyper-parameter (we used $\beta_{t,\delta}=1$ in our experiments). 

Using the notation from section \ref{sec:BayOpt}, this leads to the following model estimates:
\begin{align*}
	\MEAN(\hat f_t, x) &= \hat f_t(x)\\
	\UCB(\hat f_t, x, \delta) &= \hat f_t(x) + \beta_{t, \delta}^{1/2} \sigma_t(x)\\
	\LCB(\hat f_t, x, \delta) &= \hat f_t(x) - \beta_{t, \delta}^{1/2} \sigma_t(x)\\
	\CW(\hat f_t, x, \delta) &= 2\beta_{t, \delta}^{1/2} \sigma_t(x)
\end{align*}

The Bayesian view-point offers an alternative interpretation where the kernel estimator corresponds to the posterior mean of a Gaussian process regressor \cite{Kanagawa2018} which was exploited in deriving tighter predictive bands in \cite{Srinivas2009}.

\begin{figure}[htbp]
	\vspace{1cm}
	\includegraphics{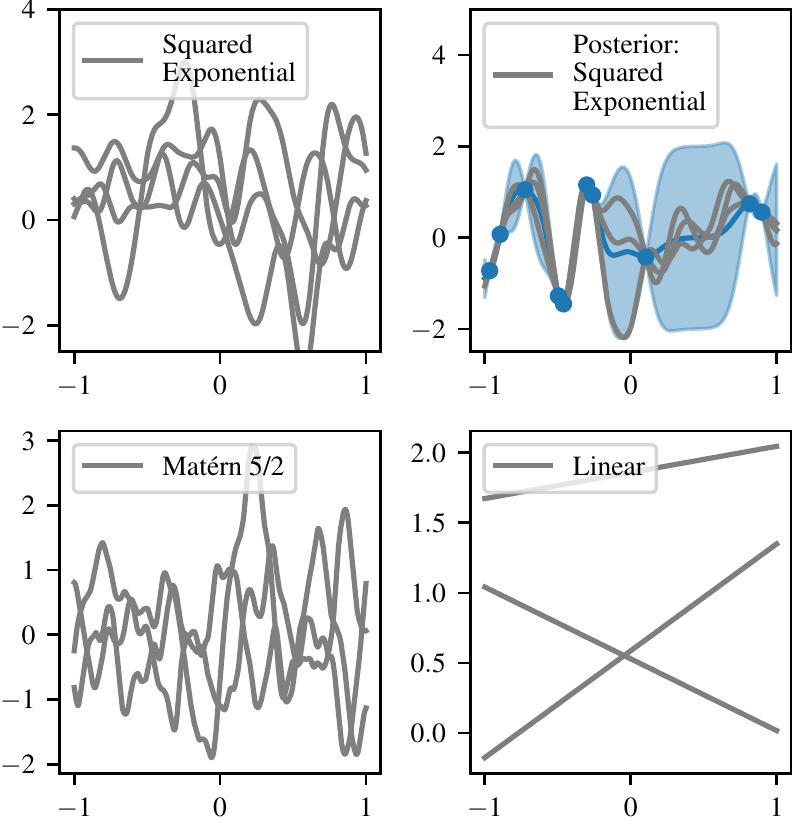}
	\caption{The plots show functions sampled from a Gaussian process defined for different kernels. In addition, the top-right plot shows the mean estimate $\hat{f}$ along with the $95\%$ confidence bands in grey. }
	\label{fig:kernels}
\end{figure}

\begin{figure*}
	\includegraphics[width=\textwidth]{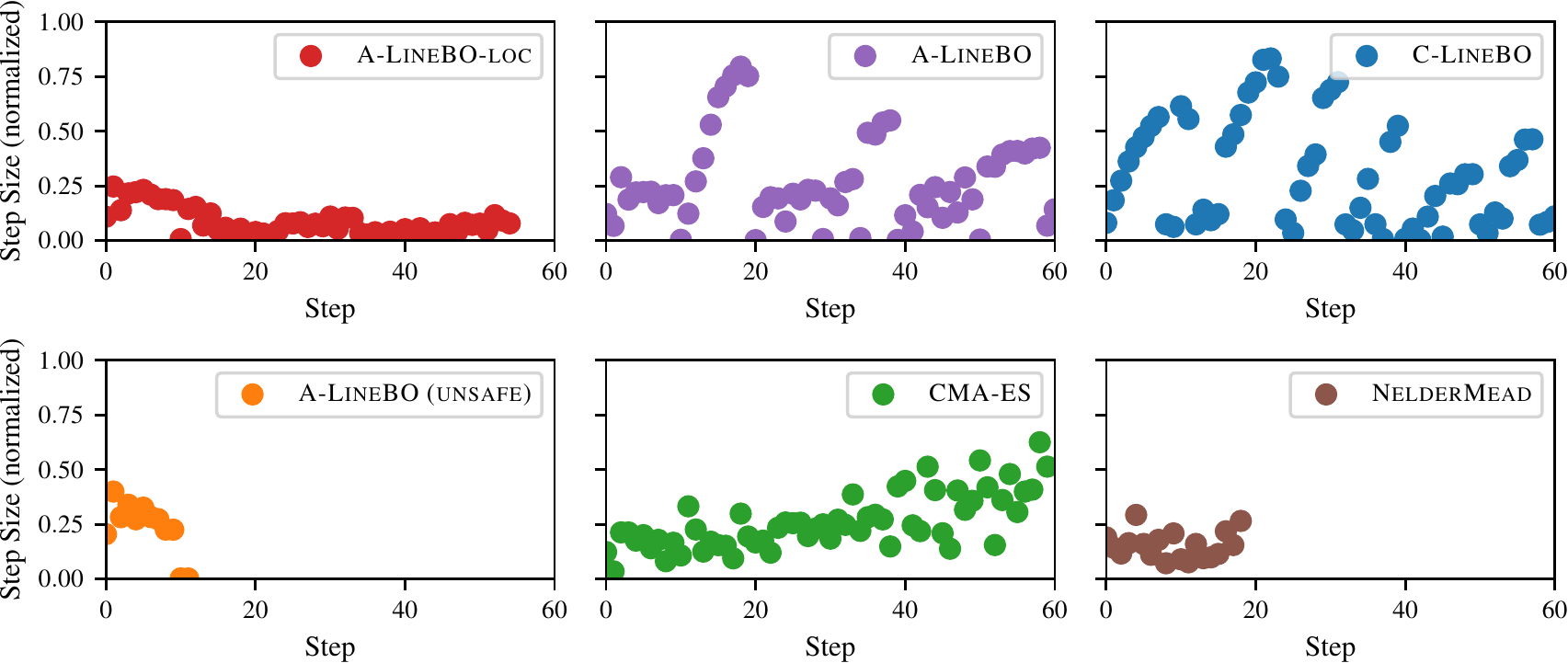}
	\caption{Each plot shows the step-size (measured in Euclidean norm) for the different methods. Note that without step-size constraints, \LineBO{} takes large steps throughout (top middle/right). We also observed that CMA-ES was slowly increasing step sizes, likely due to invariant subspaces of the objective. Unsafe \LineBO{} and \NelderMead{} stopped early because of interlocks, therefore interpretation of the data is limited.}
	\label{fig:step-sizes}
\end{figure*}

\paragraph{Kernel Choice}
The choice of the kernel is crucial for obtaining a data efficient model. Depending on the kernel choice, we can represent a different class of functions. Failure to identify the right kernel for the application leads to misspecified confidence sets and hence potential failure of the method. Different choices for the kernel are demonstrated in Figure \ref{fig:kernels} for squared exponential $k(x,y) = \exp\left(-\frac{(x-y)^2}{\gamma^2}\right)$, Mat\'ern kernel \cite{Rassmussen2004} and linear kernel $k(x,y) = x^\top y$. In the case of squared exponential kernel (as well as Mat\'ern), $\gamma$ is often referred to as lengthscale or bandwidth, and determines the smoothness of the estimated functions.

\paragraph{Bayesian optimization}
Classical Bayesian optimization \cite{Mockus1982} usually revolves around a simple procedure depicted in Figure \ref{fig:alg}, where an utility function - in this case the \textsc{UCB} - is maximized to determine the next evaluation point. Other acquisition functions are possible, see e.g. \citet{shahriari2015taking}. 

\section{Additional Experimental Data}\label{app:experiments}

Figure \ref{fig:step-sizes} shows the step-sizes for each algorithm, measured as the Euclidean distance between two consecutive input parameters on a normalized scale. Note that the step-size constrained variant takes significantly smaller steps throughout. We also found that \CMAES{} was increasing its internal step size during this run. This is due to invariant subspaces on the objective, which leads to an increased sampling (co-)variance along these directions.

\end{document}